\documentclass[twocolumn,aps,pra,showpacs,showkeys,preprintnumbers,amsmath,amssymb]{revtex4}
\usepackage{graphicx}
\usepackage{dcolumn}
\usepackage{bm}
\begin{document}
%
%
\title{Seismic Waveguide of Metamaterials}

\author{Sang-Hoon  \surname{Kim}$^{a}$} \email{shkim@mmu.ac.kr }
\author{Mukunda P. \surname{Das}$^{b}$}\email{mpd105@rsphysse.anu.au}
\affiliation{
$^a$Division of Marine Engineering, Mokpo National Maritime University,
Mokpo 530-729, R. O. Korea
\\
$^b$Department of Theoretical Physics, RSPhysSE, Institute of Advanced Studies,
The Australian National University, Canberra, ACT 0200, Australia
}
\date{\today}
\begin{abstract}
  We have developed a new method of an earthquake-resistant design
  to support conventional aseismic designs using acoustic metamaterials.
We suggest a simple and practical method to reduce the amplitude of a seismic wave
exponentially.
Our device is an attenuator of a seismic wave.
Constructing a cylindrical shell-type  waveguide
that creates a stop-band for the seismic wave,
we convert the wave into an evanescent wave for some frequency range
without touching the building we want to protect.
\end{abstract}
  \maketitle

\section{Introduction}

Earthquakes are the result of sudden release of huge amount of energy
in the Earth's crust that produces seismic waves.
A sudden forthcoming of the seismic waves with large amplitudes and low frequencies
have been great hazards to life and property.
 It is the collapse of bridges, dams, power plants, and other structures that
 causes extensive damage and loss of life during earthquakes.
Aseismic capabilities are highly relevant to public safety and
a large amount of research has gone into establishing practical analysis and
design methods for them.
  Numerous earthquakeproof engineering methods have been tried to resist earthquakes,
but still we are not so safe.

Seismic waves are a kind of inhomogeneous acoustic wave with various wavelengths.
There are two types of seismic waves: body waves and surface waves.
 P(Primary) and S(Secondary) waves are the body waves
 and R(Rayleigh) and L(Love) waves are the surface waves.
Surface waves travel slower than body waves and the amplitudes decrease exponentially with the depth.
It travels about $1 \sim 3km/sec$ with lots of variety within the depth of a wavelength \cite{vill,gion}.
The wavelengths are in the order of $100m$
and the frequencies are about $10 \sim 30Hz$,
that is, low end and just below the audible frequency.
However, they decay slower than body waves
and are most destructive because of their low frequency, long duration,
 and large amplitude.

Rayleigh waves can exist only in an homogeneous medium with a boundary and
  have transverse motion.\cite{vill,gion}
 Earthquake motions observed at the ground surface are mainly due to R waves.
  On the other hand, L waves are  polarized shear waves guided by an elastic layer.
It is this that causes horizontal shifting of the Earth during earthquakes.
L waves have both longitudinal and transverse motion and this is
 what most people feel directly during earthquakes.

Recent development of metamaterial science opens a new direction to control the seismic waves.
Farhat et al. proposed a design of a cloak to control bending waves propagated
in isotropic heterogeneous thin plates \cite{farh1,farh2,farh3}.
Their cloaking of  shear elastic waves pass smoothly into the material
rather than reflecting or scattering at the material's surface.
However, the cloaked seismic waves are still destructive
to the buildings behind the cloaked region.

In this paper we introduce a method to control seismic waves by using a
new class of materials called metamaterials. Metamaterials are
artificially engineered materials which has a special property of negative
refractive index.
We present here a solution that a metamaterial acts as
an attenuator by converting the destructive seismic wave into an
evanescent wave by making use of the imaginary velocity of stop-band of
the wave.

There are many representations to express the scale of earthquakes.
Among them the magnitude that comes from the amplitude
of the seismic waves is most important.
The common form of the magnitude, $M$, is the {\it Richter-scale}
  defined by comparing the two amplitudes in a logarithmic scale as
\begin{equation}
M=\log \frac{A}{A_o},
\label{2}
\end{equation}
where $A$ is the maximum amplitude of the seismic wave and $A_o$ is
the maximum amplitude of the background vibration and order of $\mu m$.
The equipment measures a transformed magnitude of the intensity.
Another factor of the strength is PGA(Peak Ground Acceleration)
but it is not expressed in a closed form.
We focus here on how to reduce the amplitude of the seismic wave by using
the properties of the metamaterials.

\section{Negative modulus}

Acoustic waves are created by compressibility or elasticity of the medium.
Young's modulus, $Y$, is a one-dimensional compressibility
defined by $\Delta P=Y \Delta l/l$, where $\Delta P$ is the pressure or stress
and $l$ is the length.
Shear modulus, $G$, is a two-dimensional one for a surface wave defined by
$\Delta P = G \Delta x / h$, where
$\Delta x$ is the horizontal shift and $h$ is the height of the object.
Bulk modulus, $B$, is a three-dimensional one for a body wave
defined by $\Delta P=-B \Delta V/V$.

Seismic medium of Earth crust can be considered as an accumulation
of infinite number of elastic plates.
Although the seismic surface wave is not pure two dimensional,
 the velocity is mainly dependent upon the density,
$\rho$, and shear modulus, $G$, of the seismic medium.
Seismic wave is a kind of acoustic wave
and every acoustic wave propagates following two wave equations in principle.
 Assuming the plane wave time dependence $e^{i\omega t}$,
the pressure, $p$, and the velocity, ${\vec v}$ of the wave in the two dimensions
are expressed as the Newton's 2nd law
\begin{equation}
\nabla_s p = i \omega \rho {\vec v},
\label{4}
\end {equation}
and the continuity equation
\begin{equation}
 i \omega p = G \nabla_s \cdot {\vec v},
\label{6}
\end {equation}
where $\nabla_s$ is the Laplacian operator at the surface,
  $p$ is the pressure, $\omega$ is the anguar frequency of the wave,
  and ${\vec v}$ is the velocity.

  The Eq. (\ref{4}) and Eq. (\ref{6}) generates the wave equation as
\begin{equation}
\nabla_s^2 p + \frac{\omega^2}{v^2} p = 0,
\label{8}
\end {equation}
  where the  velocity of the seismic wave is
\begin{equation}
v=\sqrt{\frac{G}{\rho}}.
\label{10}
\end{equation}
If the shear modulus becomes negative, the velocity becomes imaginary.
Then, so does the refractive index, $n$, or the inverse of the velocity as
$n=v_o/v=v_o \sqrt{\rho /G}$, where $v_o$ is the background velocity.
Since $k = 2\pi n/\lambda$, the wavevector becomes imaginary, too.
Therefore, the imaginary wavevector makes the amplitude of the seismic wave
become an evanescent wave. We call it a barrier or attenuator.
Note that the impedance $Z= \rho v = \sqrt{\rho G}$ becomes imaginary
because it is an absorption.

\begin{figure}
\resizebox{!}{0.22\textheight}{\includegraphics{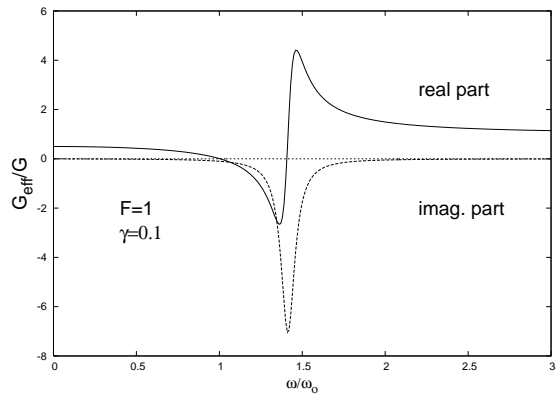}}
\caption{Real and imaginary parts of the effective shear modulus.
The negative peak of the imaginary part means an energy absorption.
$\gamma=\Gamma/\omega_o$.}
\label{fig1}
\end{figure}

Negative shear modulus of elastic media has been studied and realized very recently \cite{wu1,wu2,zhou}.
The key was the Helmholtz resonators similarly with the negative bulk modulus.
The Resonance of accumulated waves in the Helmholtz resonator
reacts against the applied pressure at some specific frequency ranges.
Then, the negative modulus is realized by passing the acoustic wave
through an array of Helmholtz resonators.
Therefore, the sound intensity decays exponentially
at some resonant frequency ranges.
In general the elastic material is described by three independent effective parameters of $G$, $B$, and $\rho$.
Therefore, sometimes the $G$ is replaced by a linear combination of $G$, $B$, and the Lam$\acute{e}$ constant \cite{vill,gion},
but it will not change the structure of the theory.
Acoustic waves from the modulus share fundamental properties of sound waves.

From the formalism of electromagnetic response in metamaterials,
effective electric permittivity and effective magnetic permeability
show negative values at some specific frequency ranges around resonances \cite{pendry}.
The Helmholtz resonator is a realization of an electrical resonance circuit by mechanical
correspondence.
It is known as that the plasmon frequency in metals or in an array of metal wires
produces the electric permittivity as \cite{caloz}
\begin{equation}
\epsilon = \epsilon_o \left[ 1- \frac{\omega_p^2}{\omega(\omega+i\Gamma)}\right],
 \label{13}
\end{equation}
where $\omega_p$ is the plasma frequency and $\Gamma$ is a loss by damping.

The Eq. (\ref{4}) is the counterpart of the Faraday's law
and Eq. (\ref{6}) is of the Ampere's law by the analogy of electromagnetism and mechanics.
The inverse of the modulus in mechanical system corresponds to the
electric permittivity in electromagnetic system.
Considering the structural loss, the general form of the effective shear modulus,
 $G_{eff}$, is given similarly with the general form of the bulk modulus as
\cite{fang1,cheng1,cheng2,lee1,lee2}
\begin{equation}
\frac{1}{G_{eff}}=\frac{1}{G}\left[ 1-\frac{F \omega_o^2}{\omega^2 -
\omega_o^2 + i \Gamma \omega} \right],
 \label{15}
\end{equation}
where $\omega_o$ is the resonance frequency and $F$ is a geometric factor \cite{caloz,ding}.
The real and imaginary part of the $G_{eff}$ is plotted in Fig. \ref{fig1}
at specific values of $F$ and $\Gamma$.
The real part can be negative at resonance  and slightly increased
frequency ranges.
The negative range of the real part is the stop-band of the wave.
When the imaginary part, the loss, is small compared with real part,
the effective shear modulus has negative value at $1 < \omega/\omega_o <
\sqrt{1+F}$.

\section{Seismic attenuator}

 We can built an attenuator or an earthquakeproof barrier of a seismic wave
  by filling-up many resonators  under the ground
  around the building that we want to protect.
  Then, the amplitude of the seismic wave that passed the waveguide is reduced exponentially
  by the imaginary wavevector at the frequency ranges of negative modulus.
 Mixing up many different kinds of resonators will cover many different corresponding
 frequency ranges of the seismic waves.

If we assume that the plain seismic wave of wavelength $\lambda$
propagates in $x-$direction, the amplitude of the wave reduces exponentially as
\begin{equation}
A e^{ikx} = A e^{ i2 \pi n x/\lambda} = A e^{-2\pi |n| x/\lambda}.
 \label{20}
\end{equation}
Let the initial seismic wave, that is, before entering the waveguide, have
amplitude $A_i$ and magnitude $M_i$,
and final seismic wave, that is after leaving the waveguide,
 have amplitude $A_f$ and magnitude $M_f$ following the Eq. (\ref{2}).
Then, $A_f$ is written as $A_i$ from Eq. (\ref{20}) as
\begin{equation}
A_i  e^{-2\pi |n| x/\lambda} = A_f.
\label{25}
\end{equation}
The amplitude of the seismic wave reduces exponentially as passing the waveguide
of metamaterials.
We can rewrite Eq. (\ref{25}) with the definition of
the magnitude in Eq. (\ref{2}) as
\begin{equation}
A_o 10^{M_i}  e^{-2\pi |n| x/\lambda} = A_o 10^{M_f}.
\label{30}
\end{equation}
Taking logarithms both sides of Eq. (\ref{30}),
we obtain the width of the waveguide, $ x \rightarrow \Delta  x$,
  as
\begin{equation}
\Delta  x = \frac{\ln 10}{2 \pi}\frac{\lambda \Delta M}{|n|}
=\frac{0.366 \lambda }{|n|}\Delta M,
 \label{35}
\end{equation}
where $\Delta M = M_i - M_f$.

\begin{figure}
\resizebox{!}{0.12\textheight}{\includegraphics{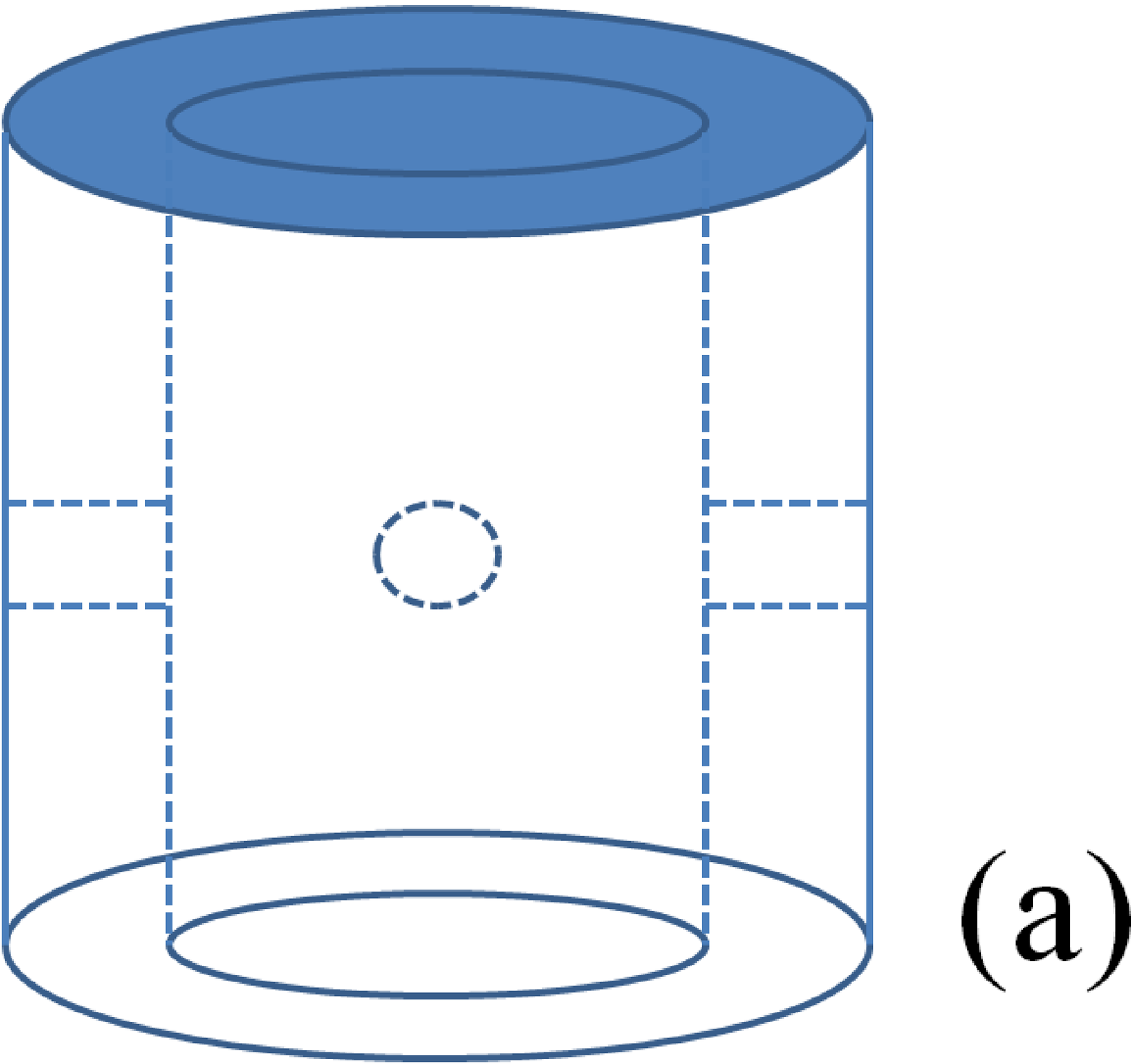}}
\resizebox{!}{0.15\textheight}{\includegraphics{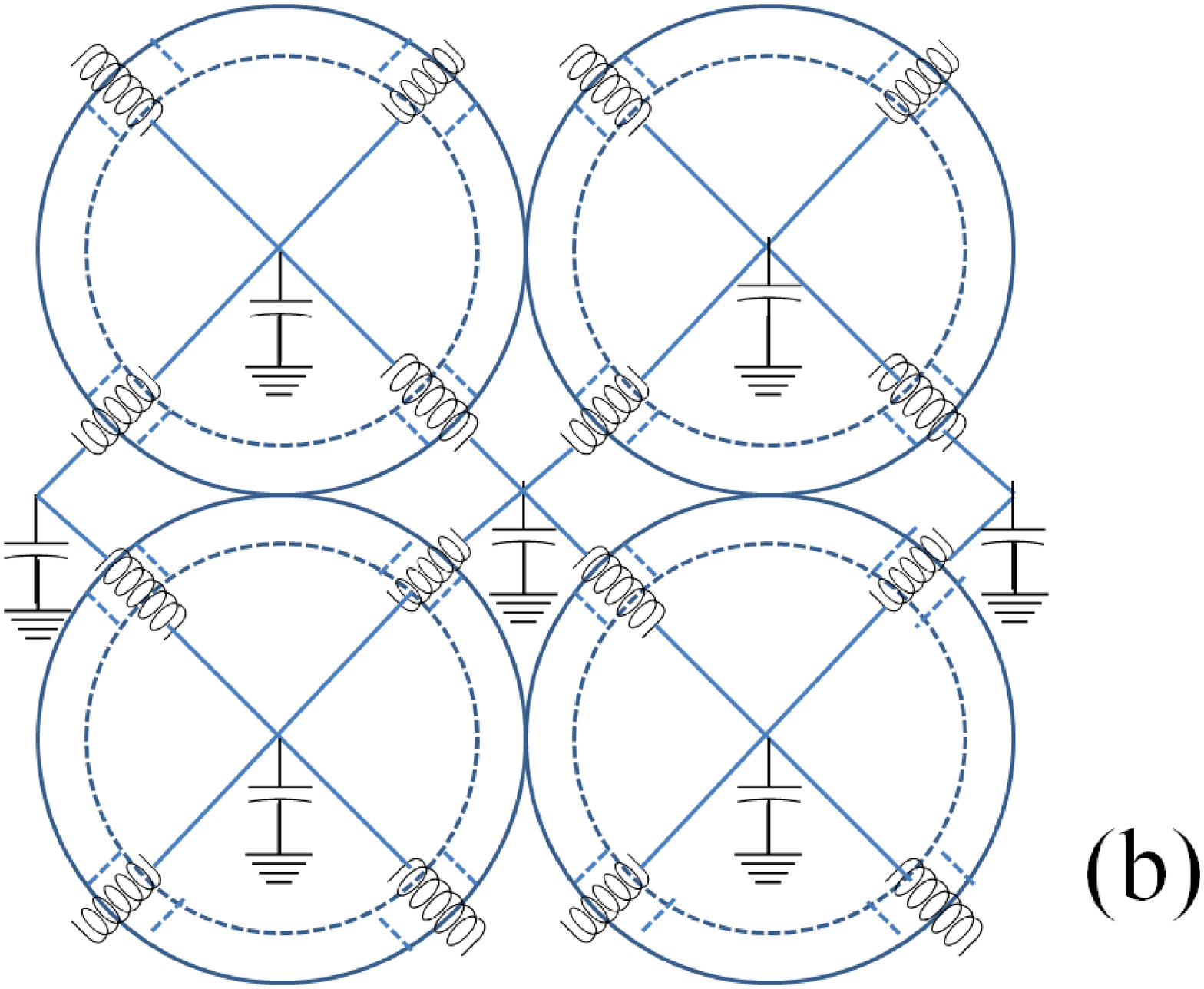}}
\caption{ (a)A sample of a meta-cylinder with 4 side holes.
The size of the cylinder is less than the wavelength of the surface waves.
(b) A combined form of the 4 meta-cylinders. An electrical analogy is shown.
 }\label{fig2}
\end{figure}

For example, if the refractive index is $n=2$
and the wavelength of the surface wave is $\lambda=100m$,
we need the waveguide of the width $\Delta x \simeq 18m$ to reduce $\Delta M=1$.
If the aseismic level of the building is $M=5$ and
the width of the waveguide surrounding the building is about $60m,$
 then the effective aseismic level of the building is increased to $M=8$.
Therefore, a high refractive index material
is desirable for a narrow waveguide.
In civil engineering earthquakeproofing methods must be practical, that is, clear to manufacture and easy to construct.
The resonator should be easy to build.
We designed an example of a resonator in Fig. \ref{fig2}.

The size of the cylinder can be estimated from the
analogy between electric circuits and mechanical pipes.
A pipe or tube with open ends corresponds to an inductor,
and a closed end corresponds to a capacitor \cite{ding,bera}.
\begin{equation}
L=\frac{\rho l'}{S}, \hspace{1cm} C=\frac{V}{\rho v^2}, \label{7}
\end{equation}
where $\rho$ is the density inside the volume, $l'$ is the effective length,
$S$ is the area of the cross-section,  $V$ is the volume,
and $v$ is the velocity inside.
From Eq. (\ref{7}) the resonant frequency is given as
\begin{equation}
\omega_o \simeq \frac{1}{\sqrt{LC}}=\sqrt{\frac{S}{l' V}}v.
\label{17}
\end{equation}
In the meta-cylinder  $l'$ is the effective length which is given by
$l' \simeq l + 0.85d$ \cite{bera},
 where $l$ is the length of the hole or thickness of the cylinder,
 and $d$ is the diameter of the hole.

 An example of the design of the meta-cylinder
  for the seismic frequency range is followings:
the diameter of the hole is order of $0.1m$,
the thickness of the cylinder is order of $0.1m$,
 and the volume inside is order of $m^3$.
Since the meta-cylinder is considerably smaller than the corresponding wavelengthes,  the array meta-cylinders behaves as a homogenized medium.

The shape of the meta-cylinder is neither necessary to be circular nor to have 4 holes.
It could be any form of a concrete box with  several side holes.
Cubic or hexagonal boxes would be fine.
Various kinds of resonators may cover various kinds of
resonance frequencies of the seismic waves.
There happens an energy dissipation of the seismic waves inside of the waveguide
and the  absorbed energy will turn into sound and heat.
They makes the temperature of the waveguide increasing
 depending on the magnitude of energy that arrives at the waveguide.

\begin{figure}
\resizebox{!}{0.2\textheight}{\includegraphics{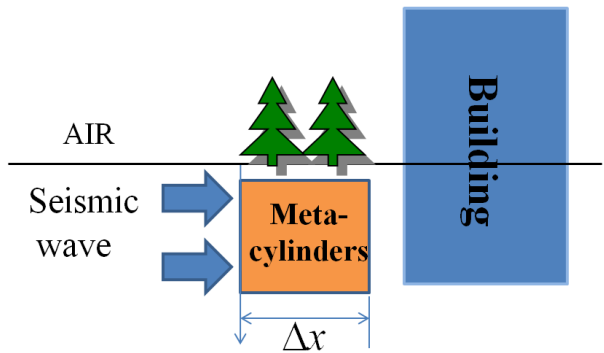}}
\caption{ A vertical landscape of the metamaterial(MTM)
barrier and the building to protect from the seismic wave. }
\label{fig4}
\end{figure}

\begin{figure}
\resizebox{!}{0.2\textheight}{\includegraphics{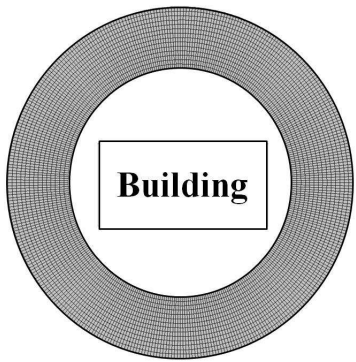}}
\caption{A sky view of a completed seismic waveguide with many meta-cylinders.}
\label{fig5}
\end{figure}

A vertical view of the metamaterial barrier with many meta-cylinders
are in Fig. \ref{fig4}.
 The width of the barrier, $\Delta x$, is predicted in Eq. (\ref{35}).
The depth of the waveguide should be at least the foundation work
of the building to protect as in Fig. \ref{fig4},
but it is not necessary to be more than the wavelength of the surface waves.
The completed form of the waveguide is the aseismic cylindrical shell of
many concentric rings in Fig. \ref{fig5}.

Seismic waves cannot pass through water. Then, we can imagine a
water barrier of a big trench filled with water. However, it is not
stable to stand a series of attacks of `foreshock $\rightarrow$ main
shock $\rightarrow$ aftershock.'
Because, if the outer part of the water trench is brought down
by a seismic wave, we do not have enough time to rebuild it to prepare for
the main shock and the after shock.
Water cannot have a high refractive index $n$ in Eq. (\ref{35}), too.
Maintaining the depth of the water trench up to the wavelength of the seismic wave
 require huge amount of water and, therefore, not be practical.


\section{Summary}

We introduced a supportive method for aseismic design.
  It is not to add another aseismic system to a building
  but to construct an earthquakeproof barrier   around the building to be protected.
   This barrier is a kind of waveguide that reduces exponentially the amplitude of the
   dangerous seismic waves.

Controlling the width and refractive index  of the waveguide,
we can upgrade the aseismic range  of the building as needed in order to defend it,
at will, without touching it.
  It could be a big advantage of the waveguide method.
This method will be effective for isolated buildings
because we need some areas to construct the aseismic shell.
It may be applicable for social overhead capitals such as power plants, dams, airports,
nuclear reactors, oil refining complexes, long-span bridges,
 express rail-roads, etc.

\section*{Acknowledgments}

This research was supported by Basic Science Research Program through
the National Research Foundation of Korea(NRF)
funded by the Ministry of Education, Science and Technology(2011-0009119).



\begin{thebibliography}{0}
 \bibitem{vill} R. Villaverde, {\it Fundamental Concepts of Earthquake Engineering}
 (CRC, New York, 2009)  Ch. 4, 5.

\bibitem{gion} V. Gioncu and F. M. Mazzolani,
{\it Earthquake Engineering for Structural Design} (Spon, New
York, 2011) p. 223.

 \bibitem{farh1} M. Farhat, S. Enoch, S. Guenneau, and A. B. Movchan,
\prl {\bf 101},  134501 (2008).

 \bibitem{farh2} M. Farhat, S. Guenneau, S. Enoch, and A. B. Movchan,
 Phys. Rev. B {\bf 79},  033102 (2009).

 \bibitem{farh3} M. Farhat, S. Guenneau, and S. Enoch,
\prl {\bf 103},  024301 (2009).

\bibitem{wu1} Y. Wu, Y. Lai,  and Z.-Q. Zhang,   Phys. Rev. B  {\bf 76}, 205313 (2007).

\bibitem{wu2} Y. Wu, Y. Lai,  and Z.-Q. Zhang,   \prl {\bf 107},  105506 (2011).

\bibitem{zhou} X. Zhou and  Z. Hu,   Phys. Rev. B {\bf 79}, 195109 (2009).

\bibitem{pendry} J. B. Pendry, A. J. Holden, D. J. Robbins, and W. J. Stewart,
 IEEE Trans. Microw. Theory tech. {\bf 47}  (1999) 2075.

\bibitem{fang1} N. Fang, X. Xi, J. Xu, M. Ambati, W. Srituravanich, C. Sun and X. Zhang,  Nature, {\bf 5},  452 (2006).

\bibitem{cheng1} Y. Cheng, J. Y. Xu, and X. J. Liu,
  \apl  {\bf 92},  051913  (2008).

\bibitem{cheng2} Y. Cheng, J. Y. Xu, and X. J. Liu,
 Phys. Rev. B {\bf 77},  045134 (2008).

\bibitem{lee1} S. H. Lee, C. M. Park, Y. M. Seo,
  Z. G. Wang, and C. K. Kim,
   J. Phys. Condensed Matter, {\bf 21},   175704 (2009).

\bibitem{lee2} S. H. Lee, C. M. Park, Y. M. Seo,
  Z. G. Wang, and C. K. Kim,   \prl {\bf 104},  054301 (2010).

\bibitem{caloz} C. Caloz and T. Itoch, {\it Electromagnetic Metamaterials} (Wiley, New York, 2006) Ch. 1.

\bibitem{ding} C. Ding, L. Hao, and X. Zhao,
 J. Appl. Phys. {\bf 108},  074911 (2010).

\bibitem{bera} L. L. Beranek, {\it Acoustics} (McGraw-Hill, New York, 1954) Ch. 3, 5.
\end{thebibliography}
\end{document}